\pdfoutput=1
\documentclass{PoS}
\usepackage{amsmath}
\title{Centre Vortices in SU(3)}

\ShortTitle{Centre Vortices in SU(3)}

\author{\speaker{Alan~\'O~Cais}\thanks{We thank the Australian Partnership for Advanced Computing (APAC) and the South Australian Partnership for
Advanced Computing (SAPAC) for generous grants of supercomputer time which have enabled this project. We also thank the University of Adelaide Faculty of Sciences for allowing us time on their Condor\texttrademark cycle-scavenging system. This work is supported by the Australian Research Council.}, Waseem~Kamleh, Ben~Lasscock, Derek~Leinweber, Lorenz~von~Smekal \\%
        
       Centre for the Subatomic Structure of Matter, School of Chemistry \& Physics,\\ The University of Adelaide, Adelaide, SA 5005, Australia\\
       E-mail: \email{alan.ocais@adelaide.edu.au}}
\author{Kurt Langfeld\\
School of Maths \& Stats, University of Plymouth,\\
Plymouth, PL4 8AA, England}

\abstract{We investigate the effectiveness of using smearing as a means to generate a preconditioning transformation for gauge fields prior to fixing to Maximal Centre Gauge. This still leaves the gauge-fixed field in the original gauge orbit. As expected, we find that this preconditioning leads to higher maxima of the gauge-fixing condition, resulting in lower numbers of P-vortices. We also find that removing vortices appears to give a loss of confinement for all cases but that the string tension as measured from vortex-only configurations drops from about 65\% to as low as 26\% when using the preconditioning method.}

\FullConference{The XXV International Symposium on Lattice Field Theory\\
                July 30 - August 4 2007\\
                Regensburg, Germany}

\begin{document}

\section{Introduction}

Quark colour confinement in hadron physics remains unexplained after more than 30 years of intense study (for a recent overview see Ref.~\cite{Alkofer:2006fu}). In lattice gauge theory two avenues of research have been most commonly adopted: confinement by $Z(N)$ centre vortices and confinement by means of abelian monopoles (for a critical discussion of both see Ref.~\cite{Greensite:2003bk}). Gauge fields can be first fixed to a particular gauge, such as Maximal Abelian Gauge (MAG) or Maximal Centre Gauge (MCG), monopoles and centre vortices are then defined by the projection of these gauge-fixed fields to $U(1)^{N-1}$ or $Z(N)$ gauge fields, respectively. Much of the progress to date has occurred in $SU(2)$ using MAG and MCG, with original findings reproducing about 90\%  \cite{Bali:1996dm} and about 100\% \cite{Bertle:2000py} of the non-abelian string tension. Removing monopole \cite{Miyamura:1995xn, Sasaki:1998ww, Bornyakov:2007fz} or centre vortex \cite{Bornyakov:2007fz,Leinweber:2006zq, de Forcrand:1999ms, Gattnar:2004gx, Gubarev:2005az} degrees of freedom from $SU(2)$ lattice gauge fields appears to leave topologically trivial, non-confining gauge fields that preserve chiral symmetry.

Work in $SU(3)$ has not progressed to this level. While initial investigations were hopeful \cite{Montero:1999by, Faber:1999sq}, subsequent results for MAG and MCG were not so encouraging \cite{Stack:2002sy, Langfeld:2003ev}, both failing to reproduce the full non-abelian string tension. Earlier studies in $SU(2)$ using MCG showed that the centre-projected configurations recovered the full string-tension, however further study into the ambiguities of the gauge-fixing procedure showed that this result is plagued by Gribov copy effects \cite{Bornyakov:2000ig, Faber:2001hq}: methods which give higher values of the gauge fixing functional produce smaller values for the vortex induced string tension. We point out that when the Laplacian Centre Gauge of Ref.~\cite{deForcrand:2000pg} (which is free of Gribov ambiguities on the lattice) is used as the fixing method the full $SU(3)$ (and $SU(2)$) string tension is recovered for the centre-projected gauge fields. However, the interpretation of this vortex matter is cumbersome in the continuum limit \cite{Langfeld:2003ev, Langfeld:2001nz}.

In this paper we focus on the Gribov problem of the $SU(3)$ centre vortex picture of confinement using the MCG gauge-fixing method. In the centre-vortex picture the gauge fields are considered to be decomposed into a long-range field $Z_\mu$ carrying all the confining fluctuations and a short-range field $V_\mu$ containing non-confining perturbations as well as other short range effects
\begin{displaymath}
  U_\mu(x)=Z_\mu(x)V_\mu(x).
\end{displaymath} 
Here $Z_\mu(x)$ is the centre element which is closest, on the $SU(3)$ group manifold, to $U_\mu(x)$. 

\subsection{Smearing as a Preconditioner}

Since the centre elements, in the centre vortex picture of confinement, correspond to the long-range physics, we employ the use of smearing, which smoothes out the short-range fluctuations, to construct a preconditioning gauge transformation for each gauge field prior to gauge-fixing \cite{Hetrick:1997yy}. 

Firstly, we smear the gauge field using any smearing algorithm (stout-link smearing \cite{Morningstar:2003gk} was used in the data shown here but other smearing algorithms are currently under investigation). We then fix the smeared field using the MCG gauge-fixing method. 

At each iteration we keep track of the total gauge transformation that has been applied to the smeared gauge field. Once the algorithm has converged we use the stored total transformation as a preconditioning gauge transformation for the unsmeared gauge field. We emphasise that the (unsmeared) preconditioned gauge field remains on the same gauge orbit since the preconditioning is merely a (specific) gauge transformation on the original links. Gauge-fixing the preconditioned field simply gives us a Gribov copy of the result from gauge-fixing the original gauge field.

\section{Identifying Vortex Matter}
 
We employ the MCG gauge-fixing algorithm as outlined in Ref.~\cite{Langfeld:2003ev}. The gauge condition we chose to maximise (with respect to the centre elements $Z_\mu(x)$) in this algorithm is 
\begin{displaymath}
V_U[\Omega] = \frac{1}{N_l}\sum_{x,\mu}\big[\frac{1}{3}\text{tr}U^{\Omega}_\mu(x)\big]\big[\frac{1}{3}\text{tr}U^{\Omega}_\mu(x)\big]^\dagger,
\end{displaymath} 
where $N_l$ is the number of links on the lattice and $U^{\Omega}$ is the gauge-transformed field.

After fixing the gauge, each link should be close to a centre element of $SU(3)$, $Z^m=e^{i\phi^m}$, $\phi^m=\frac{2\pi}{3}m$ with $m\in \{ -1,0,1 \}$.
Since, for every link, 
\begin{displaymath}
\frac{1}{3}\text{tr} U^{\Omega}_\mu(x) = u_{x,\mu}e^{i\phi_{x,\mu}} \hspace{1cm}\text{and}\hspace{1cm} \phi_{x,\mu}=\tan^{-1}\frac{\text{Im}(\text{tr} U^{\Omega}_\mu(x))}{\text{Re}(\text{tr} U^{\Omega}_\mu(x))}
\end{displaymath}
then $\phi_{x,\mu}$ should be close to some $\phi^m$, by construction of the gauge-fixing condition. We then perform the centre projection by mapping
\begin{displaymath}
SU(N)\mapsto Z_N :\hspace{1cm} U^{\Omega}_\mu(x)\mapsto Z_\mu(x) \hspace{1cm} \text{with}\hspace{1cm} Z_\mu(x)=e^{i\phi^m_{x,\mu}},
\end{displaymath}
with the appropriate choice of $\phi^m_{x,\mu}$, $m\in \{ -1,0,1 \} $.

To reveal the vortex matter we simply take a product of links around an elementary plaquette. We say a vortex pierces the plaquette if this product is a non-trivial centre element and the plaquette is then a \emph{P-vortex}. In a toy Yang-Mills model we can remove these P-vortices by hand from the configuration using $U_\mu^\prime(x)=Z^\dagger_\mu(x)U_\mu^\Omega(x)$.

\section{Results}
Calculations are performed using 100 quenched configurations with the Luscher-Weisz plaquette plus rectangle gauge action \cite{Luscher:1984xn} on a $16^3\times32$ lattice with $\beta=4.6$. Similar results are being found on 200  $20^3\times40$ lattices but since these results are currently incomplete they are not presented here.

Stout-link smearing with a smearing parameter of $0.1$ is used to construct the preconditioning transformation with the number of sweeps ranging from 0 to 12 in steps of 4 sweeps. Here, each preconditioning was conducted independently although it would be possible to use the preconditioning from an initial level to precondition subsequent levels of smearing, thereby decreasing computation time when seeking comparative results.

\begin{table}
 \hspace{-0.04\textwidth}
 \begin{minipage}{0.4\textwidth}
  \centering
  \begin{tabular}{|c||c|c|}
   \hline
   \textbf{Comparison} & \textbf{Decreased} & \textbf{Higher} \\
   \textbf{Sweeps} & \textbf{Vortices} & \textbf{Maxima} \\
   \hline
   $0\to 4$&100\%&100\%\\
   $4\to 8$&85\%&79\%\\
   $8\to 12$&79\%&64\%\\
   \hline
  \end{tabular}
  \caption{\textnormal{The percentage of configurations that have a lower number of P-vortices and the percentage of configurations that achieve a higher gauge condition maximum when comparing two different values of the preconditioning.}}
  \label{tab:transitions}
 \end{minipage}
 \hspace{ 0.033\textwidth}
 \begin{minipage}{0.5\textwidth}
  \centering
  \begin{tabular}{|c||c|c|c|}
   \hline
   \textbf{Sweeps} & \textbf{Total Iter. Blocks} & \textbf{Smear Max} & \textbf{Max}\\
    \hline
   0&$60\pm16$&****&0.7360(11)\\
   4&$89\pm15$&0.9153(16)&0.7409(9)\\
   8&$97\pm20$&0.9375(21)&0.7417(11)\\
   12&$95\pm17$&0.9461(17)&0.7421(9)\\
   \hline
  \end{tabular}
  \caption{\textnormal{For each of the sweeps used in the preconditioning: the average total (smeared gauge field fixing plus preconditioned gauge field fixing) number of iteration blocks (of 50) used, the average smeared gauge condition maximum reached and the average preconditioned gauge condition maximum reached.}}
  \label{tab:errors}
 \end{minipage}
\end{table}

If we look at the number of P-vortices located in each configuration compared with the number located on the next level of preconditioning (Table~\ref{tab:transitions}), we can see that the first level of smearing decreases the number of vortices identified in all configurations. This behaviour continues right through all levels of preconditioning although the size of the effect drops to 79\% of configurations by the last preconditioning. The magnitude of the reduction in number of P-vortices is as high as $66\pm16\%$ when using the first level of preconditioning with the effect dropping to as low about 10\% for the transition between the $8$ and $12$ sweep smearing levels. The gauge condition maximum achieved also increases in all configurations for the first level of preconditioning, with the magnitude of this effect dropping to 64\% by the final preconditioning.

We also note from Table~\ref{tab:errors} that the total number of blocks of 50 iterations of the gauge-fixing algorithm increases slightly when using the preconditioning. Typically, $\frac{2}{3}$'s of the iterations are spent fixing the smeared field and $\frac{1}{3}$ fixing the preconditioned field. The gauge-fixing maximum achieved  monotonically increases however, both for the smeared fields and for the preconditioned field (though not to the same degree) with increasing levels of smearing preconditioning. It should be noted that, regardless of the preconditioning level, the centre phases of the links of the fields always remain evenly distributed across the three possible values, reflecting the fact that the realisation of centre symmetry remains unaffected. 

\subsection{The Static Quark Anti-quark Potential}

\begin{figure}[h]
\hspace{-1cm}
$\begin{array}{c@{\hspace{1cm}}c}
\mbox{\bf Original} & \mbox{\bf Preconditioned 4 Sweeps}\\ 
\includegraphics[height=5.4cm]{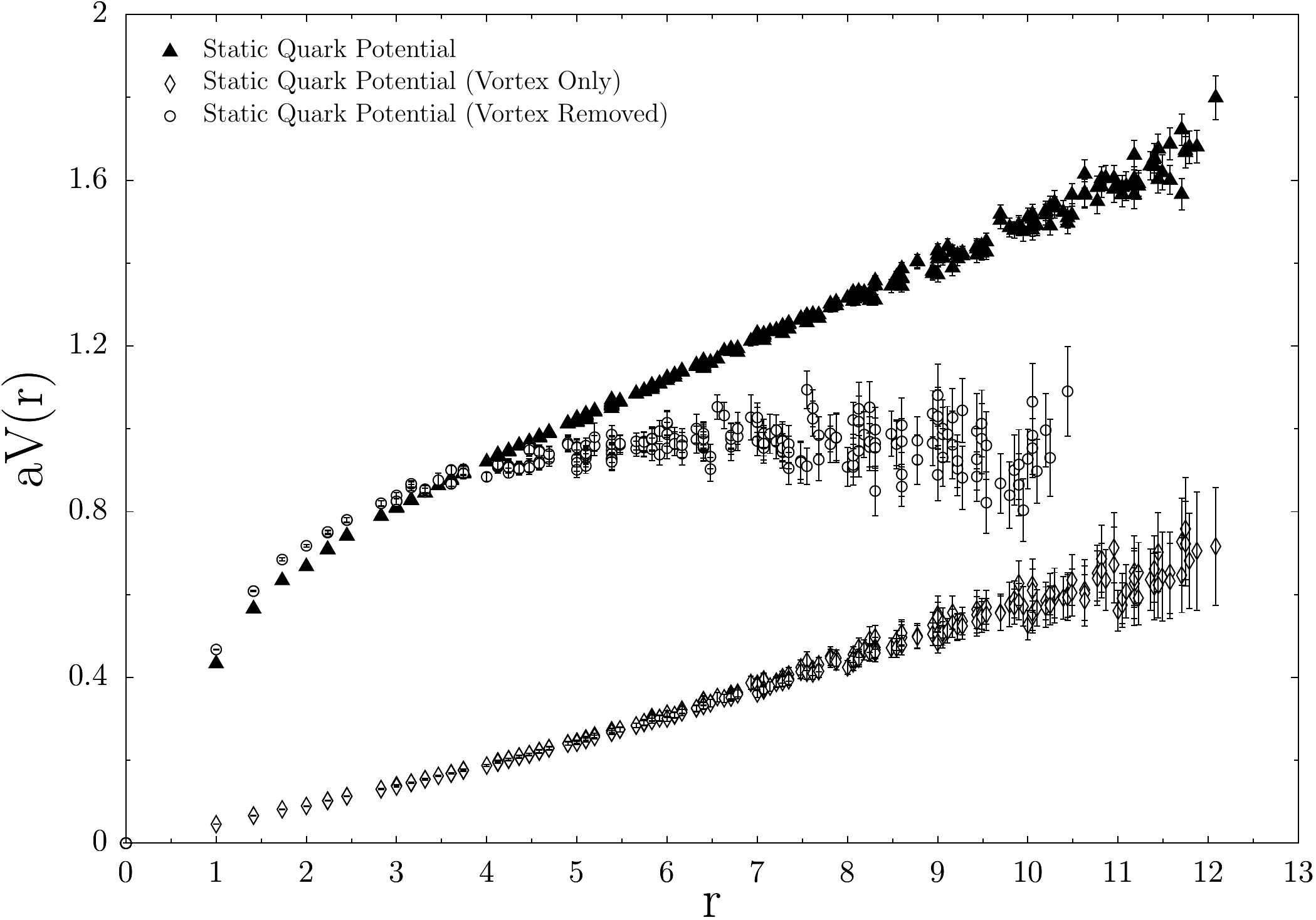} &
        \includegraphics[height=5.4cm]{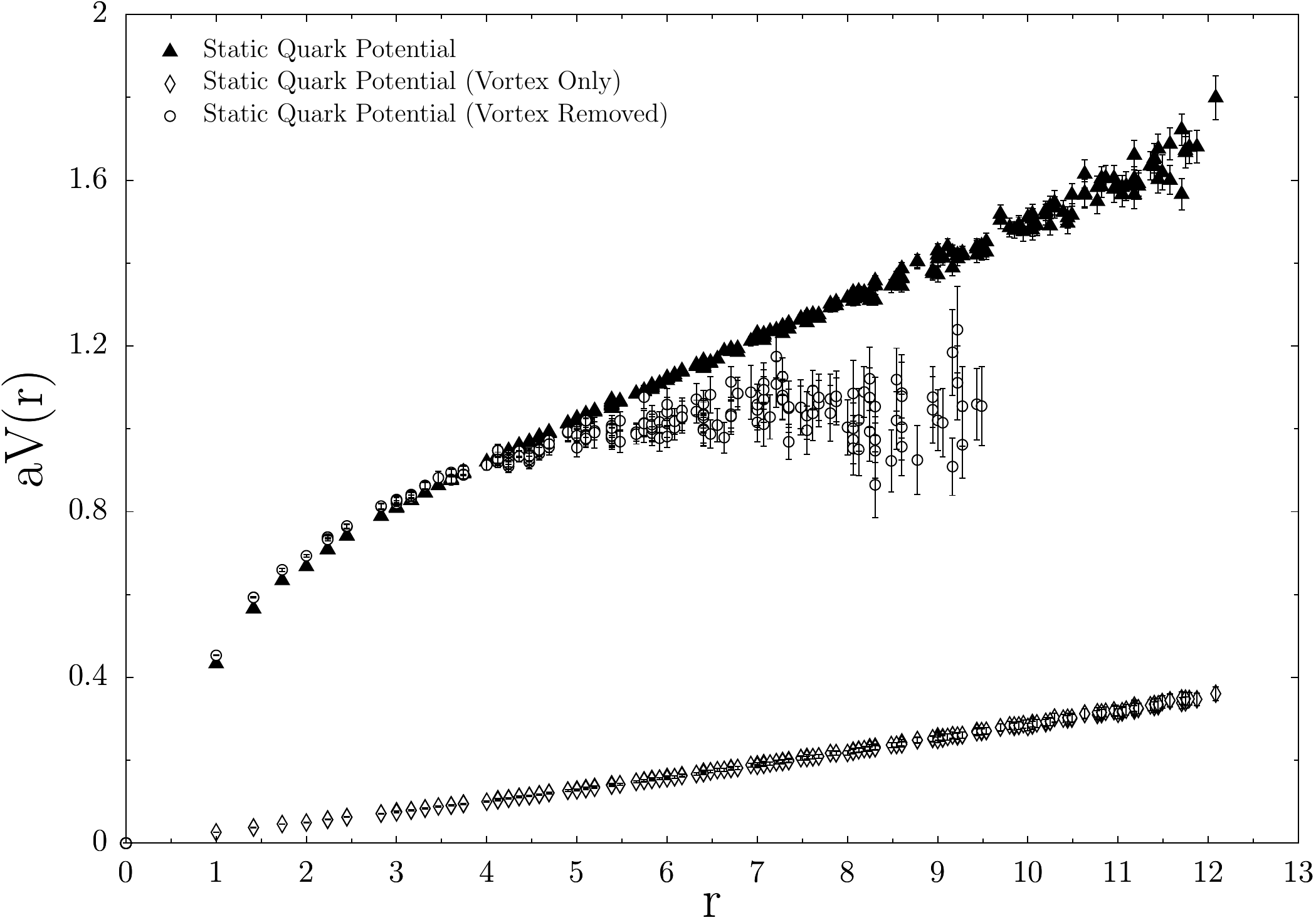} \\ [0.4cm]
\mbox{\bf Preconditioned 8 Sweeps} & \mbox{\bf Preconditioned 12 Sweeps}\\
\includegraphics[height=5.4cm]{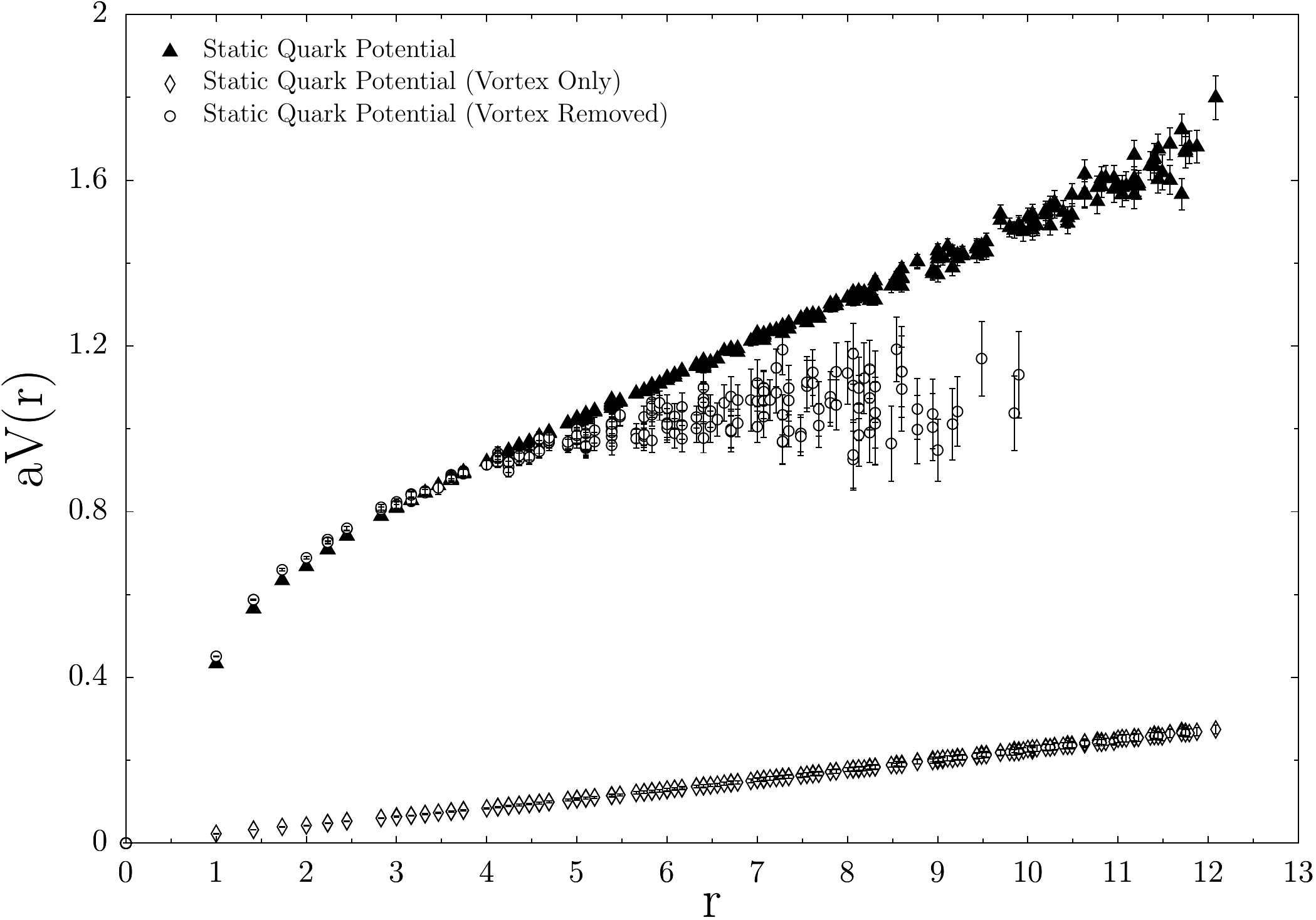} &
	\includegraphics[height=5.4cm]{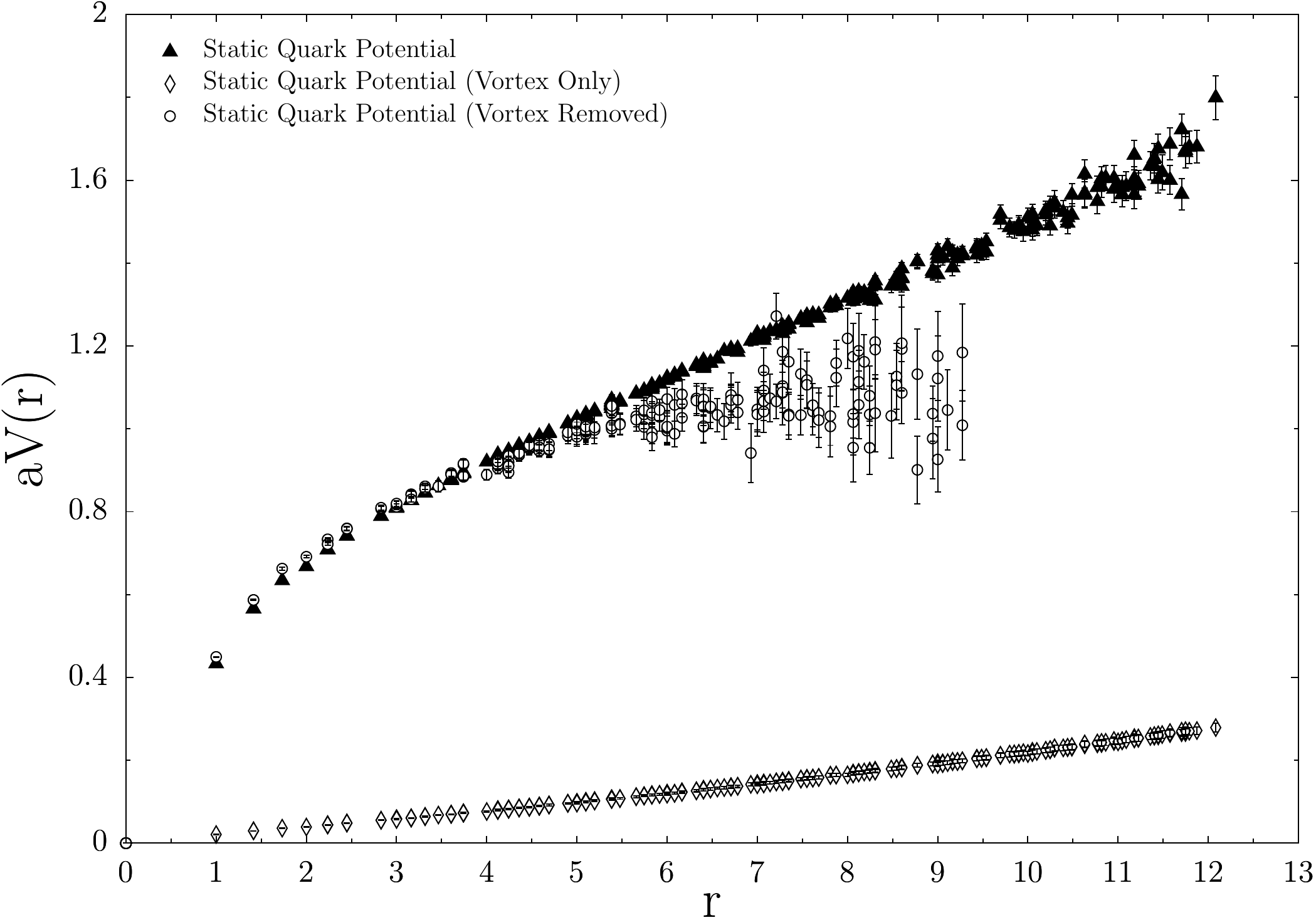} 

\end{array}$
\caption{The static quark anti-quark potential plots for each of preconditioning smearing sweeps used. Each plot contains data for the full, vortex-removed and vortex-only configurations.}
\label{figSQP}
\end{figure}

Computing the static quark anti-quark potential as a function of their separation is a two step process. Wilson loops $W(R,T)$ of extension of $R\times T$ have the large $T$ behaviour
\begin{displaymath}
\langle W(R,T)\rangle \propto \text{exp}\{ -V(r)aT\},\hspace{1cm} r:=Ra,
\end{displaymath}
 where $a$ is the lattice spacing. The method for extracting the potential is identical to that of two-point functions. To obtain the static quark anti-quark potential as a function of the quark separation we simply repeat this process for a range of values of the separation $R$. As can be seen in Figure~\ref{figSQP}, by using off-axis spatial paths for the Wilson loops, we can obtain non-integer values of $R$.

\begin{figure}[h]
\hspace{-1cm}
$\begin{array}{c@{\hspace{1cm}}c}
&\hspace{-8.7cm}{\mbox{\bf Original (Unfixed) Configuration}}  \\ 
&\hspace{-8.7cm}{\includegraphics[height=5.4cm]{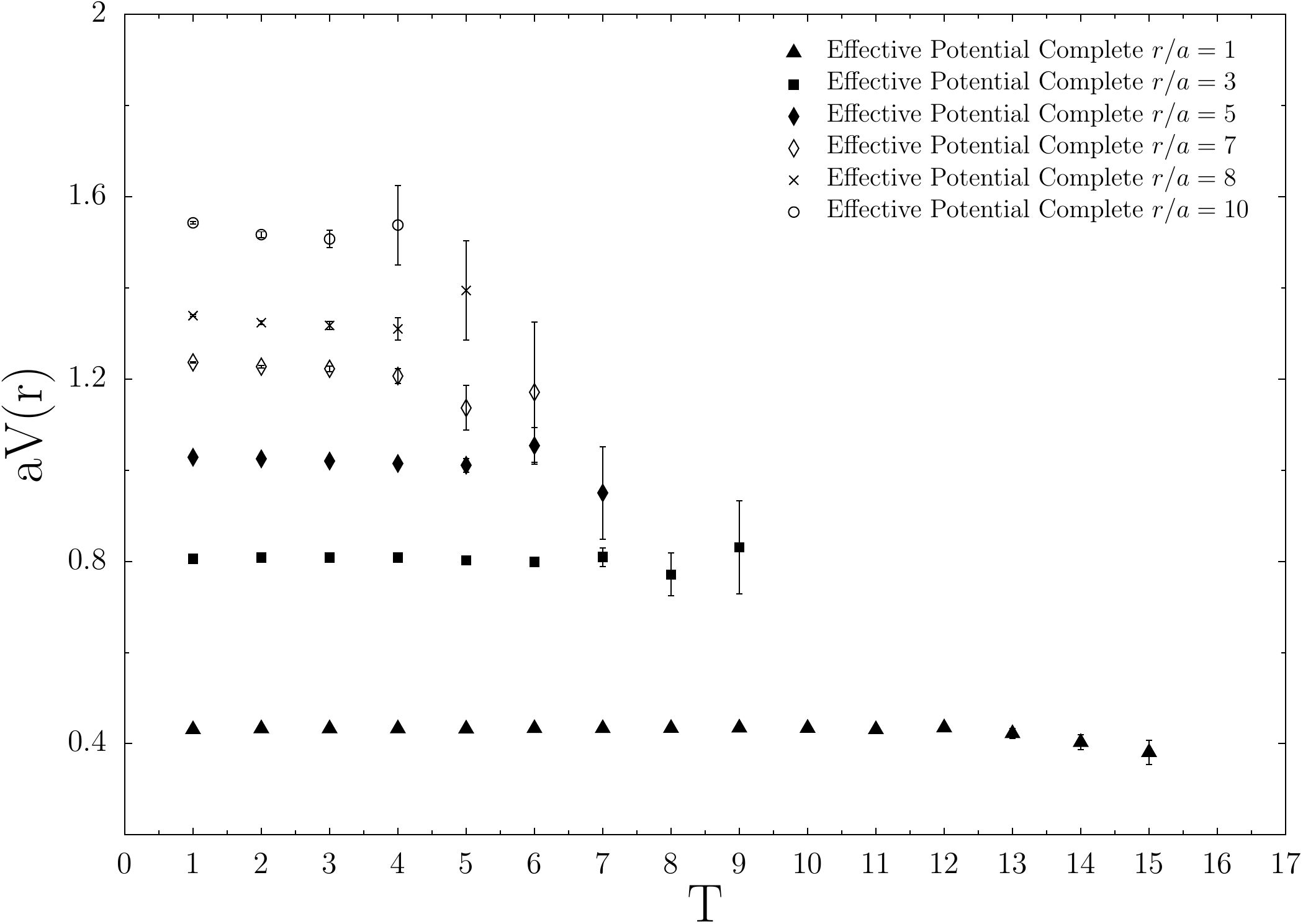}} \\[0.4cm]
\mbox{\bf Vortex Removed} & \mbox{\bf Preconditioned Vortex Removed}\\
\includegraphics[height=5.4cm]{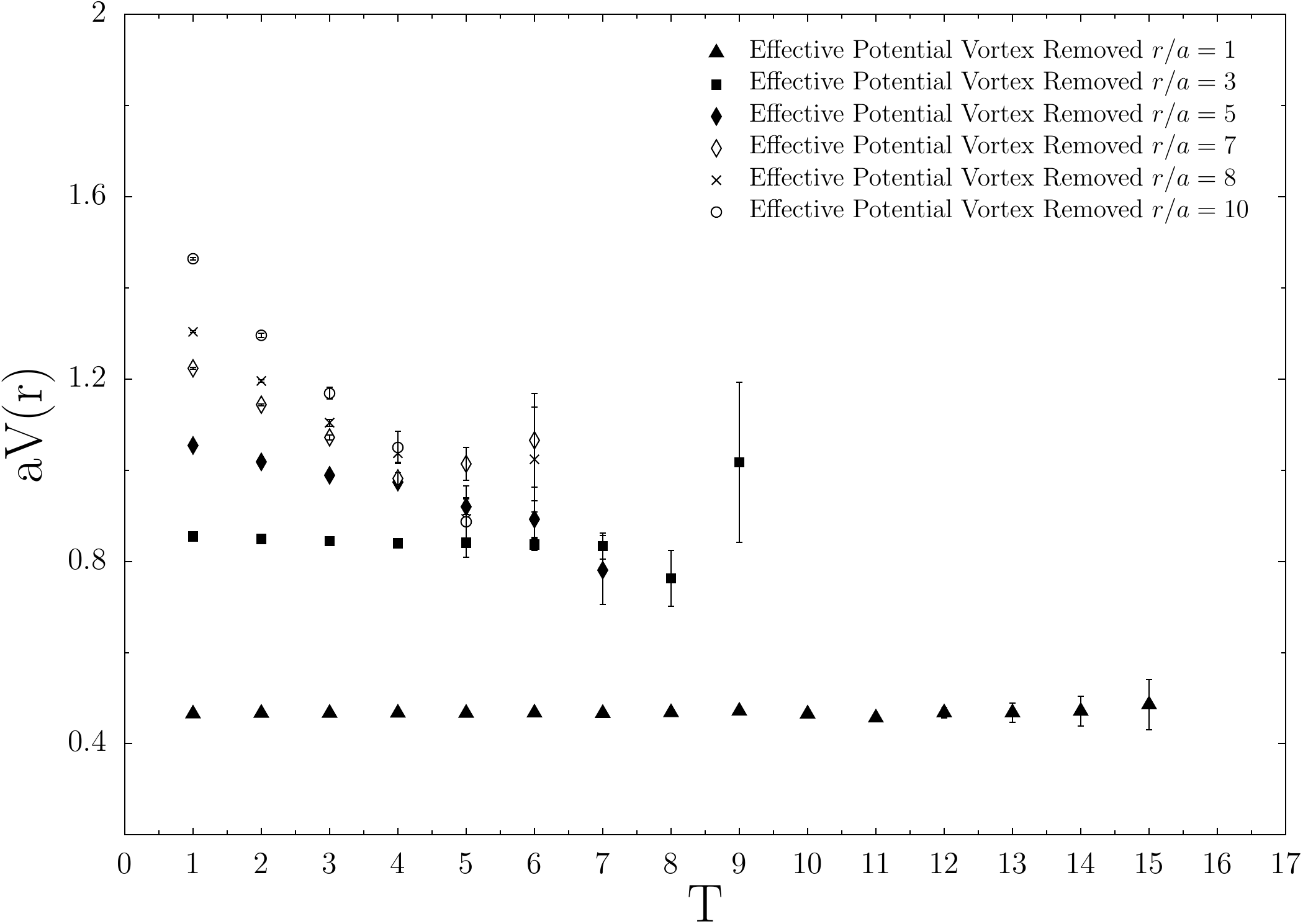} &
	\includegraphics[height=5.4cm]{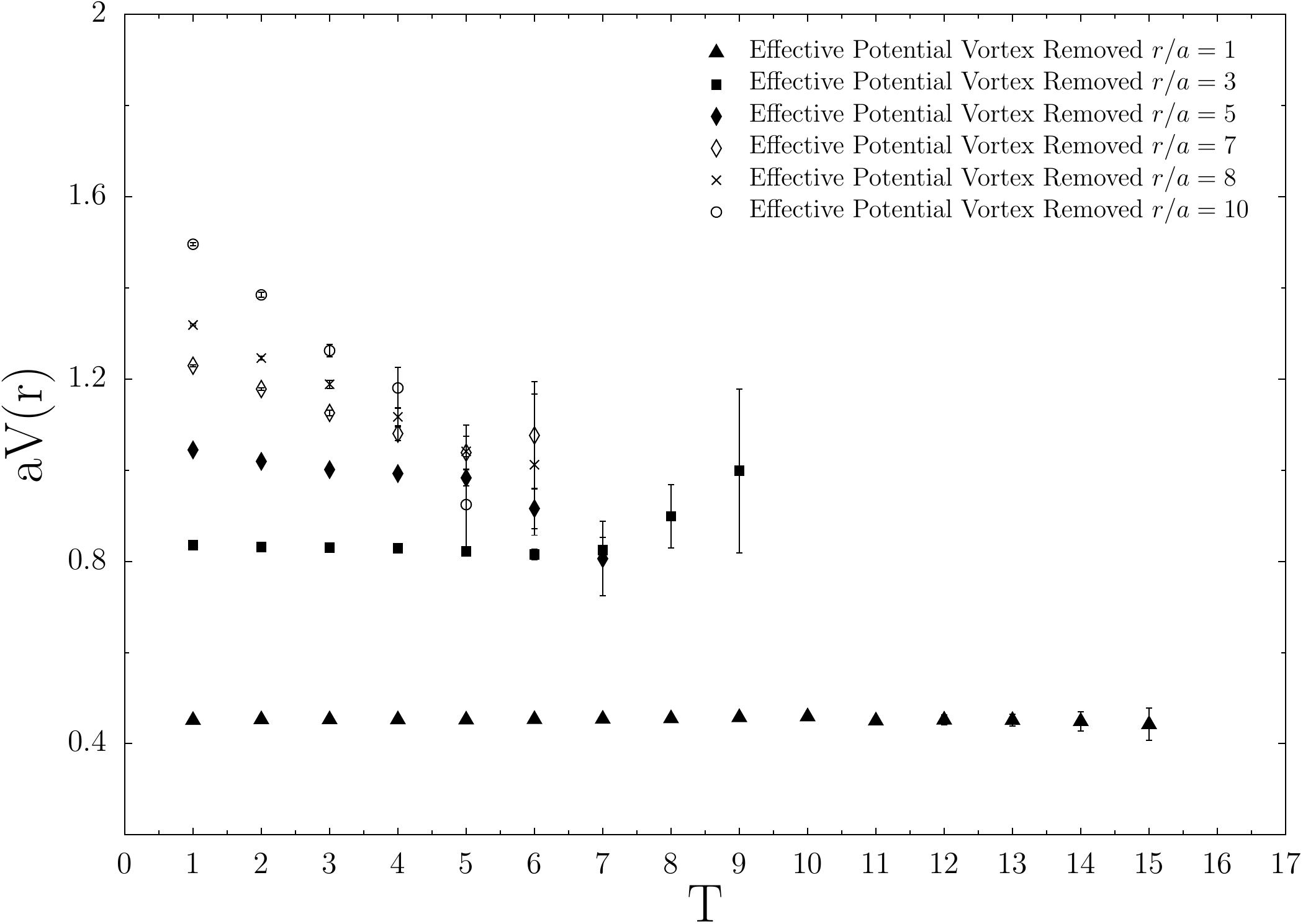} \\ [0.4cm]
\mbox{\bf Vortex Only} & \mbox{\bf Preconditioned Vortex Only}\\
\includegraphics[height=5.4cm]{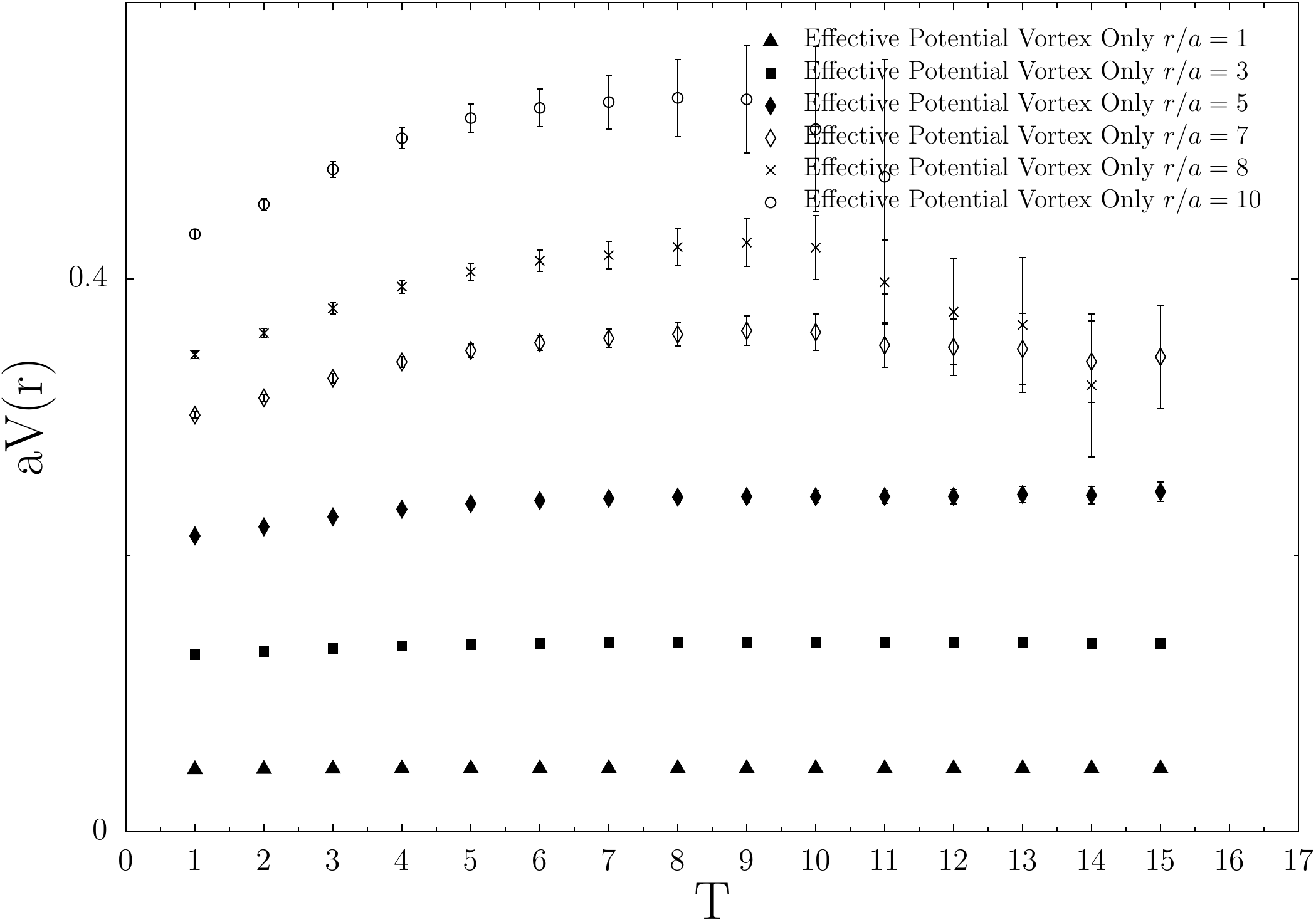} &
	\includegraphics[height=5.4cm]{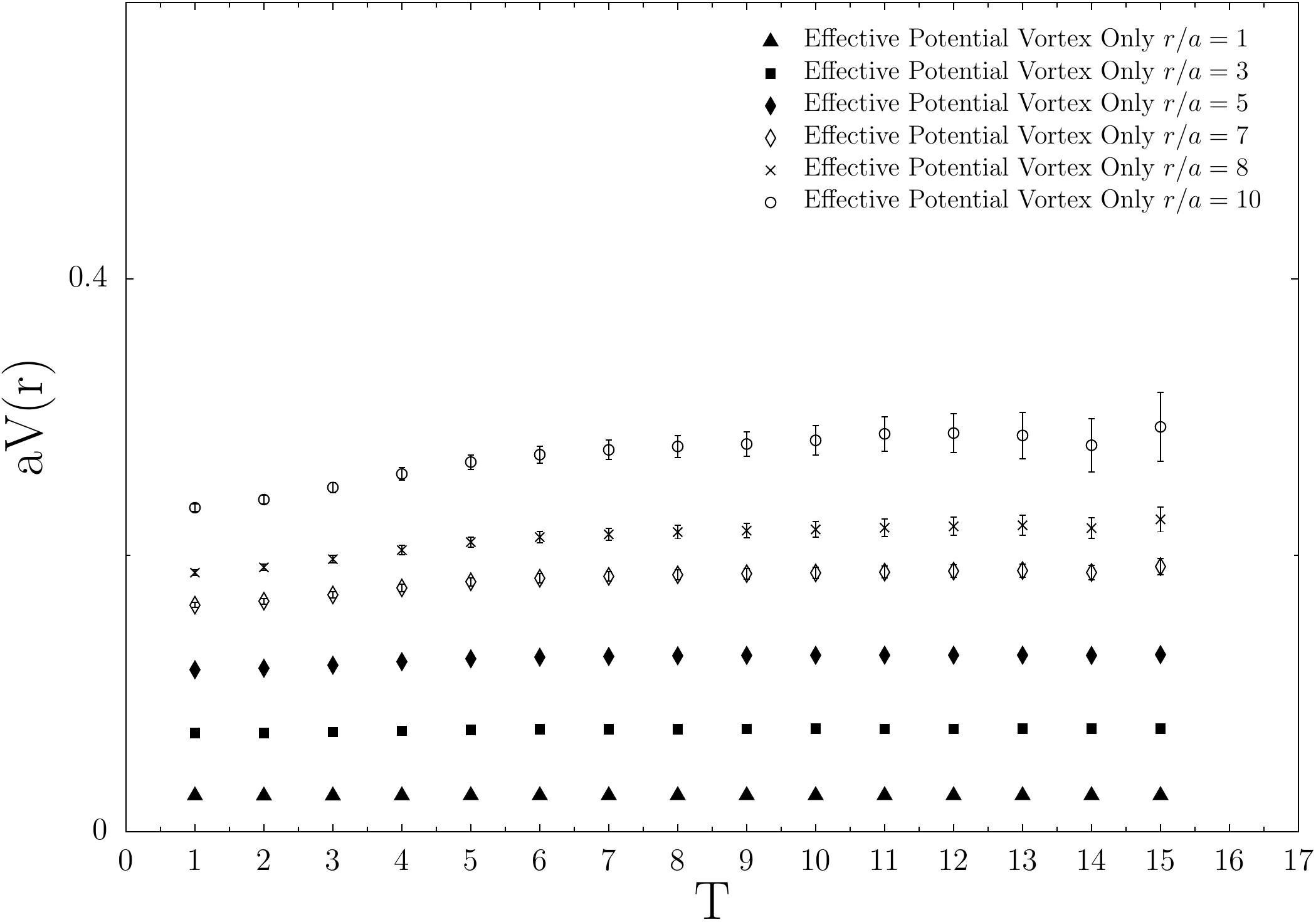}
\end{array}$
\caption{The effective potential plots for each of the untouched, vortex-removed and vortex-only configurations. On the left are the results from the unpreconditioned gauge-fixed fields and on the right the results from preconditioned gauge-fixed fields with 4 sweeps of smearing used in generating the preconditioning gauge transformation  }
\label{figeffm}
\end{figure}

In Figure~\ref{figeffm} the effective potential $V(r)$ is plotted against $T$ (by taking the log of the ratio of the value of the Wilson loops at two adjacent time slices) for two Gribov copies of each of the  vortex-only and vortex-removed configurations and comparing these to the results from the original configuration. On the left is the unpreconditioned Gribov copy and on the right is a Gribov copy that uses 4 sweeps of smearing.  We are seeking to find a plateau in the potential in each case. We note that the quality of the fit, as defined by the $\chi^2$ values, depends greatly on the fit range chosen and one must first account for the systematic drift in the data before selecting a sensible fit region. In particular, for the vortex-removed configurations, the potential does appear to become degenerate at larger values of the separation but choosing a suitable fit is challenging. We also note that the scale of the potential is greatly reduced for the vortex-only configurations while the reliability of the result increases dramatically, allowing for fits at very large $T$ values.

In Figure~\ref{figSQP} the potential $V(r)$ is plotted against the separation $r$ for our original configuration and 3 Gribov-copy ensembles. We can see that the results are consistent with the loss of confinement for the vortex-removed configurations. The preconditioned configurations better reproduce the full potential at small distances. We also note that the string tension of the vortex-only results drops dramatically when we apply our preconditioning, going from about 65\% of the full string tension to as low as about 26\%.

\section{Conclusions} 

We have shown that using smearing as a preconditioner to the MCG gauge fixing algorithm increases the maximum of the gauge-fixing condition. We find that these higher maxima correspond to a lower number of measured of P-vortices. When looking at the static quark anti-quark potential for the vortex-only and vortex-removed configurations we find results consistent with loss of confinement for all the vortex-removed configurations. We also find that the preconditioning decreases the measured string tension of the vortex-only results, reducing them from 65\% to as low as 26\% of the full string tension.

Similar to what has been observed in $SU(2)$ \cite{Bornyakov:2000ig,Faber:2001hq}, there appears to be a significant anti-correlation between the value achieved in the gauge fixing functional and the percentage string tension reproduced by centre vortices. Although the fundamental modular region of MCG would be an ideal candidate for a unique definition of vortex texture, it seems that the vortex matter arising from the first Gribov region as a whole has a bigger phenomenological relevance. Further work is continuing on larger lattice volumes. Results to date are consistent with findings here.

\end{document}